\documentstyle[12pt]{article}
\begin{document}\noindent\thispagestyle{empty}
\unitlength1cm\begin{center}
{\large \bf

QUANTUM CORRECTIONS TO THE ENTROPY\\[8pt]
 FOR HIGHER SPIN FIELDS IN HYPERBOLIC SPACE}
\vspace{1cm}

M. Bordag and A.A. Bytsenko\footnote{He leaves from: State
Technical University, 195251 St.Petersburg, Russia}

Institute for Theoretical Physics, Leipzig
University\\Augustusplatz 10, 04109 Leipzig, Germany

\end{center}

gr-qc/9412054

NTZ -23-94

\vspace{1cm}
\section*{Abstract}
We calculate the one-loop corrections to the free energy and to
the entropy for fields with arbitrary spins in the space
$S^1\otimes H^N$.  For conformally invariant fields by means of
a conformal transformation of the metric the results are valid
in Rindler space with $D=N+1$ dimensions. We use the zeta
regularization technique which yields an ultraviolet finite
result for the entropy per unit area. The problem of the
infinite area factor in the entropy which arises equally in
Rindler space and in the black hole background is addressed in
the light of a factor space $H^N/\Gamma$.
\newpage 
\section{Introduction}

Quantum fields in black hole backgrounds have been actively
investigated during the last years \cite{1}, but only recently
an interest to thermodynamic properties and in particular to the
entropy have been renewed \cite{2}. Several attemps have been
undertaken to calculate quantum corrections to the thermodynamic
entropy in spaces with horizons. A large number of works have
been carried out in the limit of infinite black hole mass when
the Schwarzschild spherical horizon surface becomes planar and
the metric becomes the Rindler metric.

It is known that the black hole entropy stored in quanta near
the horizon is ultraviolet divergent. In Rindler space several
authors have found such a divergence
\cite{3,4,5,6}. This is in contradiction with the finiteness of the Bekenstein-
Hawking entropy \cite{7}.

Recently the statistical mechanics mode counting was carried out
for the entropy spectrum of scalars in Rindler space with the
help of WKB approximation
\cite{4}. The free energy for a massless field in D=4 dimensional spacetime, in
accordance with previous results, comes out proportional to
$\beta^{-4}$ ($\beta$ is the inverse temperature) and to the
area of the horizon. It has been shown that the entropy per unit
area is quadratically divergent near the horizon.  Since the
level density diverges due to the infinite shift of frequencies,
a cutoff parameter should be introduced \cite{4,5,8}. Such
quantum corrections to the density of entropy, as has been
pointed out in \cite{4}, are equivalent to the quantum
corrections to the gravitational coupling. Therefore, the
entropy divergencies obtained by state counting are closely
related to the conventional ultraviolet divergencies of
canonical quantum gravity.

An independent calculation of the finite temperature stress
energy tensor gaves the same results \cite{5}.  An interesting
method to compute the entropy mostly for $D=2$ was developed in
\cite{6}.
In two dimensions the divergence is logarithmic and the
coefficients are cutoff independent. For $D>2$ the heat kernel
of the Laplace-Beltrami operator in Rindler space have been
used.  As a result, the free energy lower integration limit
needs a short distance regularization.  The heat kernel
techniques in computing quantum corrections in Rindler space
have been used in
\cite{8}, where
the computation was done in geometry with different topology
(without the conical singularity).

In this paper we suggest a powerful method for the computation
of the first quantum correction to the free energy associated
with fields of arbitrary spins on the manifold $M=S^1\otimes
H^N$. Such a manifold can be obtained in the result of a
conformal transformation of the Euclidean Rindler space.  In
Sect. 2 we discuss the general technique based on the zeta
regularization approach for the calculation of the one-loop free
energy.
We show in Sect. 3 that the ultraviolet divergencies of the free
energy are removed by the zeta-regularization approach.  The
free energy per unit area is represented by a series in inverse
powers of $\beta$ (i.e., as a high temperature expansion) whoose
coefficients are cutoff independent. This series has been
analytically continued to all $\beta$. In the conclusions a
comparison of the results is done with the corresponding
corrections obtained for the conformal invariant and minimally
coupled fields.  We discuss the problem of regularizing the
divergencies in the entropy which result from the infinite
horizon area.

\section{The Zeta Functional Regularization of the Free Energy}

Let us start with the functional integration of the partition
function associated with Rindler space. The geometry of the
$D$-dimensional Euclidean Rindler space can be written as
follows
\begin{equation}
{\rm d}s^2=\xi^2{\rm d}\tau^2+{\rm
d}\xi^2+\sum_{i=1}^{N-1}{\rm d}y_i^2\,.
\label{1}\end{equation}
Here $y_i$ are the $N-1$ transverse flat coordinates ($D=N+1$),
$\tau$ is the Euclidean time (periodically identified with
period $\beta$) and the lines $\xi={\rm const}$ correspond to
uniformly accelerated observers.  It is convenient to use the
optical metric \cite{9} $\overline{g}=gg_{00}^{-1}=g\xi^{-2}$ in
order to define the appropriate functional integration
\cite{8,10}
\begin{equation}
Z[\overline{g},\beta ]=\int D[\Phi]~{\rm
e}^{-S[\Phi]}\,,\label{2}\end{equation} where $S[\Phi]$ is the
Euclidean action related to the quantum field $\Phi$ in the
$D$-dimensional manifold of the form $M=S^1\otimes H^N$ and
$H^N$ is the simply connected real hyperbolic space.

It should be noted that the measure of the path integral
(\ref{2}) is formally regularized with respect to some inner
product of the fields. Generally speaking, the inner product is
ill-defined at the horizon (see, for example,
\cite{10}). But actually such a choice of the measure is suited to obtain
the same thermodynamics as the one associated with Rindler mode
counting \cite{4,8,11,12}.

The partition function can be given by
\begin{equation}{\rm log}Z^\pm [\overline{g},\beta ]=
\pm\frac{1}{2}{\rm log}{\rm det}A^\pm(\beta)\,,\label{3}\end{equation}
where the sign (+) (resp. (-)) refer to Dirac spinor (resp.
integer spin-s fields), $A^\pm(\beta)=\partial_\tau^2+L^\pm_N$,
where $L^\pm_N$ is the self-adjoint Laplace-Beltrami operator
acting in the space $H^N$.  The eigenvalues of the operator
$A^\pm(\beta)$ have the form
\begin{equation}
\omega^{(s)}_n=\left[{2\pi (n+s)\over\beta}\right]^2+\lambda^2,~~n\in Z\,.
\label{4}\end{equation}
Here $\lambda^2$ are the eigenvalues of the operator $L_N$.

The generalized (Riemann) zeta function related to the operator
$A(\beta)$ can be represented using the Mellin transform of the
heat kernel $K(t\!\!\mid\!\! A(\beta))={\rm Tr}\exp
(-t(A(\beta))$:
\begin{equation}\zeta(x\mid A^\pm(\beta))={1\over \Gamma (s)}
\int_0^\infty{\rm d}t~t^{x-1}K(t\mid A^\pm(\beta))\,.
\label{5}\end{equation}
Note that the ''global'' function $\zeta(x\!\!\mid\!\!
A(\beta))$ depends on $x\in M$ only, since the manifold $M$ is a
homogeneous space.

Using the relations
\[\sum_{n=-\infty}^\infty {\rm e}^{- n^2\beta^2/4t}=
2{\frac{\sqrt{\pi t}}{\beta}}
\sum_{n=-\infty}^\infty {\rm e}^{-4pi^2n^2t/\beta^2}
\]
which are identities for the Jacobi's elliptic theta functions
$\Theta_3(v,q)=\sum_{n=-\infty}^\infty q^{n^2}{\rm e}^{{\rm
i}2\pi n v}$ and $\Theta_4(v,q)=\sum_{n=-\infty}^\infty (-1)^n
q^{n^2}{\rm e}^{{\rm i}2\pi n v}$, one can rewrite $\zeta(x\mid
A (\beta))$ in the form (see \cite{13} for more details)
\begin{eqnarray}
\zeta(x\mid A^\pm (\beta))&=&\sum_{n=-\infty}^\infty\zeta(x\mid
L^\pm_N+[2\pi(n+s)/\beta]^2)\nonumber\\
&=&{\beta \Gamma (x-1/2)\over 2\sqrt{\pi}\Gamma
(x)}\zeta(x-1/2\mid L^\pm_N)+\nonumber\\ &&+{\beta\over
2\sqrt{\pi} \Gamma (x)}\int_{0}^\infty{\rm d}t~t^{x-3/2}K(t\mid
L^\pm_N)\left(\Theta^\pm (\beta ,t)-1\right)\,,
\label{6}\end{eqnarray}
where
\begin{equation}\Theta^\pm(\beta ,t)=\left\{ {
\Theta_3(0,{\rm e}^{-\beta^2/4t}),~s=0,1,2, ...
\atop
\Theta_4(0,{\rm e}^{-\beta^2/4t}),~s=1/2~~~~.       }\right.
\label{7}\end{equation}

A complex integral representation for the zeta function can be
obtained with the help of Mellin-Parseval identity
\begin{equation}\int_0^\infty{\rm d}t~f(t)~g(t)~=
{}~{1\over 2\pi {\rm i}}\int_{\Re
z=\sigma}\widehat{f}(z)~\widehat{g}(1-z)~{\rm d}z\,,
\label{8}\end{equation}
where $\widehat{f}(z)$ ($\widehat{g}(z)$) is the Mellin
transform of $f(t)$ ($g(t)$), $t^{z-1}f(t)\in L(0,\infty )$ and
$\sigma$ being a real number in the strip in which
$\widehat{f}(z)$ and $\widehat{g}(1-z)$ are analytic. Let us
suppose that $f(t)=t^{z-3/2}K(t\mid L^\pm_N)$,
$g(t)=(-1)^{2sm}\exp (-n^2\beta^2/4t)$.  Then, using Eq.
(\ref{6}) and the Mellin-Parceval identity (\ref{8}) one can
obtain
\begin{eqnarray}
\zeta(x\mid A^\pm(\beta))&=&
{\beta \Gamma (x-1/2)\over 2\sqrt{\pi}\Gamma (x)}
\zeta(x-1/2\mid L^\pm_N)\nonumber\\
&&+{1\over \sqrt{\pi}\Gamma (x) 2\pi {\rm i}} \int_{\Re z=c}{\rm
d}z~\zeta^\pm(z)\Gamma ({z\over 2})\Gamma ({z-1\over 2}+x)
\nonumber \\
&&\hspace{3cm}\times\zeta ({z-1\over 2}+x\mid
L^\pm_N)\left({\beta\over 2}\right)^{-(z-1)}\,,
\label{9}\end{eqnarray}
where $\zeta^-(z)=\zeta_R(z)$ is the Riemann zeta function,
$\zeta^{+}(z)=(1-2^{1-z})\zeta_R(z)$ and $c>N+1$.

Let us point
out the contribution $Z[\overline{g},\beta ]$ of conformally
invariant fields to the partition function $Z[g,\beta ]$ in the
Rindler space.  The renormalized free energy ${\cal F}[g,\beta
]$ for two conformally related static spaces with metrics
$\overline{g}_{\mu\nu}={\rm e}^{-2\omega}g_{\mu\nu}$ can be
rewritten in the form \cite{14}
\begin{equation}{\cal F}^\pm [g,\beta ]=
\pm\beta^{-1}\log Z^\pm[g,\beta ]=
{\cal F}^\pm [\overline{g},\beta ]+\Delta{\cal F}^\pm [\omega,g
]\,.
\label{10}\end{equation}
We may apply further Eq. (\ref{10}) to the particular case when
$\omega =\frac{1}{2}\log \xi^2$ and $\overline{g}_{\mu\nu}={\rm
e}^{-2\omega}g_{\mu\nu}$ is an ultrastatic metric. It should be
noted that the term $\Delta{\cal F}^\pm [\omega,g ]$ for two
conformally related theories is proportional to $\beta$ and,
hence, does not contribute to the entropy. So, we will suppress
this term below.

{}From Eq.(\ref{9}) one can obtain the representation for the free
energy which is related to the free energy by
means of equation
\begin{equation}
{\cal F}^\pm[\overline{g},\beta ]
=\mp\frac{1}{2}\beta^{-1}\zeta^\prime (0\mid A^\pm (\beta))\,.
\label{101}\end{equation}
As a result, we have
\begin{eqnarray}
{\cal F}^\pm[\overline{g},\beta ]&=&\mp\frac{1}{2}
\zeta^{(r)}(-1/2\mid L^\pm_N)\pm {1\over 2\pi{\rm i}}
\int_{\Re z=c}{\rm d}z~
\zeta^\pm (z)\Gamma (z-1)\zeta ({z-1\over 2}\mid L^\pm_N)\beta^{-z} \nonumber\\
&\equiv &F_0^\pm+F^\pm (\beta )\,.\label{11}\end{eqnarray} Here
$F_0^\pm$ is the vacuum energy, $F^\pm(\beta)$ is the temperature
dependent part of ${\cal F}^\pm [\overline{g},\beta ]$, and we
have introduced the notation
\begin{equation}
\zeta^{(r)}(-1/2\mid L^\pm_N)= {\rm FP}~\zeta(-1/2\mid L^\pm_N)+
(2-2\log 2)~{\rm Res}_{z=-1/2}~\zeta (z\mid L^\pm_N)\,.
\label{12}\end{equation}
In Eq. (\ref{12}) the symbols FP and Res denote the finite part
and residue of the function at the specified point, respectively
(for more details see \cite{13}).  As it has been pointed out in
\cite{14,13}, the contribution to the free energy of the fermionic
field $F^+(\beta )$ can be obtained from the bosonic part of
$F^-(\beta )$. This relation can be easily reproduced from Eq.
(\ref{11}) and the result is
\[
F^+(\beta )=2F^-(2\beta )-F^-(\beta )\,.\]

\section{Quantum Corrections Associated with Arbitrary Spin
Fields in $S^1\otimes H^N$}

For a noncompact rank one symmetric space $H^N$ (the rank $M$ is
the dimension of the commutative algebra of invariant
differential operators, Laplace operators for instance) the
related zeta function can be constructed by the help of the
spectral function $\mu (\lambda )$ known as the Plancherel
measure \cite{15}. In the case of a Riemann noncompact symmetric
space with negative curvature the explicit form of $\mu (\lambda
)$ is given by
\[\mu (\lambda )=[C(\lambda )C(-\lambda )]^{-1}\,,\]
where $C(\lambda )$ is the Harish-Chandra function \cite{16,13}.
It can be given in terms of a product over the positive roots of
the symmetric space.  The spectral function is essential in the
construction of the zeta function in a noncompact space. It
takes the form
\begin{equation}
\zeta(z\mid L^\pm_N)=\int_0^\infty{{\rm d}\lambda~\mu(\lambda )\over
(\lambda^2+C_s^2)^z}\label{14}\end{equation}
and $C_s$ are known constants depending on the mass of the
fields and on the curvature $R=-N(N-1)a^{-2}$.  We choose the
radius $a=1$ and in the final results the dependence on $a$ can
be restored.

Here, we add a remark concerning the role of the coefficients $C_s$. As it is
known,  for a conformal invariant massless scalar field it holds $C_0=0$ and
the corresponding Lap|lace-Beltrami operator has no gap in its spectrum.  In
that case we keep the coefficient $C_0\ne 0$ until the end of the calculations
and use it as a regularization parameter.  In this way we get a well defiend
zeta function which is suited for analytical continuation.

The Plancharel measure for spin-$s$ fields in $H^N$ has been
calculated recently in \cite{16,17}. It reads
\begin{equation}
\mu^\pm(\lambda ,s)={\pi\lambda\over [2^{N-2}\Gamma (N/2)]^2}\tanh
[\pi(\lambda +{\rm i}s)]~~\sigma^\pm(\lambda
,s)\label{15}\end{equation} with \[\tanh [\pi(\lambda +{\rm
i}s)]=\left\{ {\tanh (\pi\lambda ),~~s=0,1,...\atop \coth
(\pi\lambda ),~~s=1/2}\right.\] and
\[\sigma^-(\lambda ,s)\hspace{14cm}\]
\begin{equation}=\left\{ {
\left[\lambda^2+\left(s+{N-3\over
2}\right)^2\right]\prod_{j=0}^{q-2}~\left(\lambda^2+j^2\right)\equiv\sum_{k=1}^q~a^{(s)}_{k,N}\lambda^{2k},~~N=2q+1,
\atop
\left[\lambda^2+\left(s+{N-3\over
2}\right)^2\right]\prod_{j=1/2}^{q-5/2}~\left(\lambda^2+j^2\right)\equiv\sum_{k=0}^{q-1}~b^{(s)}_{k,N}\lambda^{2k},~~N=2q
, }\right.\label{16}\end{equation}
\[\sigma^+(\lambda ,1/2)\hspace{14cm}\]
\begin{equation}=\left\{ {
\left[\lambda \coth
(\pi\lambda)\right]^{-1}\prod_{j=1/2}^{q-1/2}~\left(\lambda^2+j^2\right)
\equiv  \left[\lambda \coth (\pi\lambda)\right]^{-1}
\sum_{k=0}^q~a^{+}_{k,N}\lambda^{2k} ,
{}~~N=2q+1,
\atop
\prod_{j=1}^{q-1}~\left(\lambda^2+j^2\right)\equiv\sum_{k=0}^{q-1}~b^{+}_{k,N}\lambda^{2k},~~N=2q
, \hspace{6cm}}\right.\label{17}\end{equation} where $q\in {\cal Z}_+$.
The coefficients $a^{(s)}_{k,N}$, $b^{(s)}_{k,N}$, $a^+_{k,N}$,
$b^+_{k,N}$ are defined by expanding the products into
polynomials in $\lambda^2$ in Eqs. (\ref{16}) and (\ref{17}).
In Eq. (\ref{16}), for $N=3$, the product is to be omitted and
$a^{(s)}_{0,3}=s^2$, $a^{(s)}_{1,3}=1$.  For $N=4$, the product
is also omitted and we have $b^{(s)}_{0,4}=(s+1/2)^2$,
$b^{(s)}_{1,4}=1$ (the spectral functions on $H^2$ for spin 0
and 1 are both given by $\mu^-(\lambda ,s)=\pi \lambda\tanh
(\pi\lambda )$, $b^+_{0,2}=1$).

Finally, the zeta function can be written as \cite{17,18}
\[\zeta(z\mid L^-_N)\hspace{14cm}\]
\begin{equation}=\left\{
\begin{array}{l}
\frac{1}{2}
g(s)A(N)\sum_{k=1}^qa^{(s)}_{k,N}C_s^{2k-2z+1}B(k+1/2,z-k-1/2)~,~~N=2q+1\\[12pt]
\frac{1}{2} g(s)A(N)\sum_{k=0}^{q-1}b^{(s)}_{k,N}\left[
C_s^{2k-2z+2}B(k+1,z-k-1)-4I^-_k(C_s,z)\right]~,~~N=2q\end{array}\right.
\label{18}\end{equation}

\[\zeta(z\mid L^+_N)\hspace{14cm}\]
\begin{equation}=\left\{
\begin{array}{l}
2^{[N/2]-1}A(N)\sum_{k=0}^qa^{+}_{k,N}C_{1/2}^{2k-2z+1}B(k+1/2,z-k-1/2)~,~~N=2q+1\\[12pt]
2^{N/2-1}A(N)\sum_{k=0}^{q-1}b^{+}_{k,N}\left[
C_{1/2}^{2k-2z+2}B(k+1,z-k-1)+4I^+_k(C_{1/2},z)\right]~,~~N=2q\end{array}\right.
\label{19}\end{equation}
where $B(x,y)=\Gamma (x)\Gamma (y)/\Gamma (x+y)$ is Euler's beta
function,
\begin{equation}
I^\pm_k(C_,z)=\int_0^\infty{ {\rm
d}\lambda~\lambda^{2k+1}(\lambda^2+C^2)^{-z}
\over
{\rm e}^{2\pi\lambda}\mp 1}\,,\label{20}\end{equation}
\begin{equation}g(s)={(2s+N-3)(s+N-4)\over (N-3)!s!},~~~A(N)={A\over
2^{N-1}\pi^{N/2}\Gamma (N/2)}\,,
\label{21}\end{equation}
and $A$ is the area of the manifold $M$. For $N=3$ we should
take $g(0)=1$ and $g(s)=2$ for $s\ge 1$.

For integer spins and odd $N$, the zeta function (\ref{18}) is
meromorphic in the complex $z$-plane with simple poles at
$z=N/2~, N/2-1, ...$ and exhibits trivial zeros at
$z=0,-1,-2,...$.  For even $N$, the integral term in (\ref{18})
is analytic in $z$, but the first term carries a finite number
of first order poles at $z=N/2~, N/2-1,~ ...~ ,1$. Finally, the
spinor zeta function (\ref{19}) has first order poles at the
same points (at $z=N/2~, N/2-1, ...$ for odd $N$, and at
$z=N/2~, N/2-1,~ ...~,1$ for even $N$).

In order to obtain the Laurent series representation for the
statistical sum it is convenient to use the Mellin-Barnes
representation (\ref{11}).  Using the zeta functions related to
the operators $L^\pm_N$ given by eqs. (\ref{18}) and (\ref{19})
we obtain
\begin{equation}
F^\pm(\beta )=\pm\frac{1}{2\pi{\rm i}}\int_{\Re z=c_0}{\rm
d}z~\varphi^\pm (z,N)\,, ~~~~~~~~(c_0>N/2)\,.
\label{22}\end{equation}
Here we introduced the notations
\[
\varphi^-(z,2q+1)\hspace{14cm}\]
\begin{equation}=-\frac{g(s)A(2q+1)}{4\sqrt{\pi}}
\sum_{k=1}^{q}~a^{(s)}_{k,N}C^{2k+2}_s\Gamma (k+\frac{1}{2})\Gamma
(z+\frac{1}{2})\zeta_R(2z+1)\Gamma (z-k-\frac{1}{2})\left({C_s\beta\over
2}\right)^{-(2z+1)}\,,
\label{23}\end{equation}

\[
\varphi^-(z,2q)\hspace{14cm}\]
\[=-\frac{g(s)A(2q)}{4\sqrt{\pi}}
\sum_{k=0}^{q-1}~b^{(s)}_{k,N}C^{2k+3}_s\Gamma (k+1)\Gamma
(z+\frac{1}{2})\zeta_R(2z+1)\Gamma (z-k-1)\left({C_s\beta\over
2}\right)^{-(2z+1)}\]
\begin{equation}\hspace*{2cm}+4g(s)A(2q)\sum_{k=0}^{q-1}b^{(s)}_{k,N}\zeta_R(2z+1)\Gamma (2z)I^-_k(C_s,z)\beta^{-(2z+1)}
\,,
\label{24}\end{equation}

\[
\varphi^+(z,2q+1)\hspace{14cm}\]
\begin{equation}=\frac{2^{q-2}A(2q+1)}{\sqrt{\pi}}
\sum_{k=0}^{q}~a^{+}_{k,N}C^{2k+2}_{1/2}\Gamma (k+\frac{1}{2})\Gamma
(z+\frac{1}{2})\zeta^+(2z+1)\Gamma (z-k-\frac{1}{2})\left({C_{1/2}\beta\over
2}\right)^{-(2z+1)}\,,
\label{25}\end{equation}

\[
\varphi^+(z,2q)\hspace{14cm}\]
\[=\frac{2^{q-2}A(2q)}{\sqrt{\pi}}
\sum_{k=0}^{q-1}~b^{+}_{k,N}C^{2k+3}_{1/2}\Gamma (k+1)\Gamma
(z+\frac{1}{2})\zeta^+(2z+1)\Gamma (z-k-1)\left({C_{1/2}\beta\over
2}\right)^{-(2z+1)}\]
\begin{equation}\hspace*{2cm}+2^{q+2}A(2q)\sum_{k=0}^{q-1}b^{+}_{k,N}\zeta^+(2z+1)\Gamma (2z)I^+_k(C_{1/2},z)\beta^{-(2z+1)}
\,.
\label{26}\end{equation}

For odd (even) dimension $N=2q+1$ ($N=2q$), the function
$\varphi^-(z,N)$ is meromorphic in $z$. It has first order poles
at $z=0, ~z=N/2,N/2-1,N/2-2,...$ ($z=-1/2,~z=
N/2,N/2-1,N/2-2,...$) and one second order pole at $z=-1/2$
($z=0$).  The function $\varphi^+(z,N)$ has the same properties
with the only difference that for even $N$ the poles at
$z=0,-1/2$ are simple ones, while for odd N the pole at $z=-1/2$
is of second order and there is no pole at $z=0$.

Now it is useful to move the integration contour in (\ref{22})
to the left up to infinity. Thereby it will cross all the poles
just mentioned which results in contributions from the
corresponding residues. We obtain the following series
representation:
\begin{eqnarray}
F^-(\beta )_{2q+1} &=& -\frac{g(s)A(2q+1)}{4\sqrt{\pi}}
\sum_{k=1}^{q}~a^{(s)}_{k,N}C^{2k+2}_s\Gamma (k+\frac{1}{2})\nonumber \\
&&\hspace*{1cm}\times\left\{
\sum_{j=0}^k(-1)^j\zeta_R(2k-2j+2){\Gamma (k-j+1)\over\Gamma
(j+1)}\left({C_s\beta\over 2}\right)^{-(2k-2j+2)}
\right. \nonumber
\\
&&\hspace*{1cm}+{ (-1)^{k+1}\pi^{3/2}\over\Gamma (k+3/2)}
\left({C_s\beta\over 2}\right)^{-1}
+{ (-1)^{k+1}\over\Gamma (k+2)}\left[\gamma+\log
\left({C_s\beta\over 4\pi}\right)\right]\nonumber \\
&&\left.
\hspace*{1cm}+\Xi^{\rm -}_{k,2q+1} \left({C_s\beta\over 2\pi}\right)
\right\}\,,\nonumber \\ && \label{27}\end{eqnarray}

\begin{eqnarray}
F^-(\beta )_{2q} &=& -\frac{g(s)A(2q)}{4\sqrt{\pi}}
\sum_{k=0}^{q-1}~b^{(s)}_{k,N}C^{2k+3}_s\Gamma (k+1)\nonumber \\
&&\hspace*{1cm}\times\left\{
\sum_{j=0}^k(-1)^j\zeta_R(2k-2j+3){\Gamma (k-j+\frac{3}{2})\over\Gamma
(j+1)}\left({C_s\beta\over 2}\right)^{-(2k-2j+3)}
\right.           \nonumber \\
&&\hspace*{1cm}+{ (-1)^{k}\pi^{1/2}\over\Gamma (k+2)} \log
\left(C_s\beta\right)\left({C_s\beta\over 2}\right)^{-1}+
{(-1)^{k+1}\pi\over 2\Gamma (k+5/2)}
\left.+\Xi^{\rm -}_{k,2q} \left({C_s\beta\over 2\pi}\right)\right\}\nonumber \\
&&\hspace*{-1cm}+g(s)A(2q)\sum_{k=0}^{q-1}~b^{(s)}_{k,N}\left\{
\left[\frac{\rm d}{{\rm d}z}I^-(k,z)_{\mid_{z=0}}+{(-1)^k(1-2^{-2k-1})\over
4(k+1)}\left(\gamma -\log \beta^2 \right)
B_{2k+2}\right]\beta^{-1}\right.\nonumber \\
&&\hspace*{3cm}\left.-2
I^-(k,-\frac{1}{2})+\Psi^-\left({\beta \over 2\pi}\right)\right\}\,,
\label{28}\end{eqnarray}
\begin{eqnarray}
F^+(\beta )_{2q+1} &=& \frac{2^{q-2}A(2q+1)}{\sqrt{\pi}}
\sum_{k=0}^{q}~a^{+}_{k,N}C^{2k+2}_{1/2}\Gamma (k+\frac{1}{2})\nonumber \\
&&\hspace*{1cm}\times\left\{
\sum_{j=0}^k(-1)^j\zeta^+(2k-2j+2){\Gamma (k-j+1)\over\Gamma
(j+1)}\left({C_{1/2}\beta\over 2}\right)^{-(2k-2j+2)}
\right. \nonumber
\\
&&\hspace*{1cm}+{ (-1)^{k}\over\Gamma (k+2)}\left[\gamma+\log
\left({C_{1/2}\beta\over \pi}\right)\right]\left.
+\Xi^{\rm +}_{k2q+1}\left({C_{1/2}\beta\over 2 \pi}\right)
\right\}\,,\nonumber \\ && \label{29}\end{eqnarray}
\begin{eqnarray}
F^+(\beta )_{2q} &=&
\frac{2^{q-2}A(2q)}{\sqrt{\pi}}
\sum_{k=0}^{q-1}~b^{+}_{k,N}C^{2k+3}_{1/2}\Gamma (k+1)\nonumber \\
&&\hspace*{1cm}\left\{
\sum_{j=0}^k(-1)^j\zeta^+(2k-2j+3){\Gamma (k-j+\frac{3}{2})\over\Gamma
(j+1)}\left({C_{1/2}\beta\over 2}\right)^{-(2k-2j+3)}
\right.           \nonumber \\
&&\hspace*{1cm}+{ (-1)^{k+1}\pi\over\Gamma (k+2)} \log
2~\left({C_{1/2}\beta\over 2}\right)^{-1}+ {(-1)^{k}\pi\over
2\Gamma (k+5/2)}
\nonumber \\
&&\hspace*{1cm}\left.+
\Xi^{\rm +}_{k,2q}\left({C_{1/2}\beta\over 2 \pi}\right)
\right\}\nonumber \\
&&+2^{q}A(2q)\sum_{k=0}^{q-1}~b^{+}_{k,N}\left\{
\frac{(-1)^k}{2(k+1)}\log 2~ B_{2k+2}~\beta^{-1}\right.\,\label{30} \\
&&\hspace*{1cm}\left.+2
I^+_k(C_{1/2},-\frac{1}{2})+\Psi^+\left({\beta\over 2\pi}\right)\right\}\,,
\nonumber\end{eqnarray}
with
\begin{equation}\Xi^{\rm \pm}_{k,2q+1} \left(X\right)=
\frac{1}{\sqrt{\pi}}\sum_{j=k+2}^\infty(-1)^j\zeta^\pm(2j-2k-1){\Gamma
(j-k-\frac{1}{2})\over\Gamma (j+1)}X^{2j-2k-2}\,,
\label{31a}\end{equation}
\begin{equation}\Xi^{\rm \pm}_{k,2q} \left(X\right)=
\frac{1}{\sqrt{\pi}}
\sum_{j=2+k}^\infty(-1)^j\zeta^\pm(2j-2k-2){\Gamma (j-k-1)\over\Gamma
(j+1)}X^{2j-2k-3}\,,
\label{32a}\end{equation}
\begin{equation}\Psi^-(X)=\frac{1}{\pi}\sum_{j=1}^\infty{(-1)^j\over
j}\zeta_R(2j)I^-_k(C_s,-j)\left(X\right)^{2j-1}\,,\label{psim}\end{equation}
\begin{equation}\Psi^+(X)=\frac{2}{\pi}\sum_{j=1}^\infty{(-1)^j\over
j}(1-2^{2j})\zeta_R(2j)I^+_k(C_{1/2},-j)\left(X\right)^{2j-1}\,\label{psip}\end{equation}
and $\gamma$ is the Euler-Masceroni constant, $B_{2k}$ are the
Bernoulli numbers. Eqs. (\ref{27}) - (\ref{30}) give a series
respesentation for the free energy which is very convenient for
high temperature expansion.  Let us remark, that the
regularization has been already removed in these expressions. They are finite.
That means the pole at $s=0$ is absent, as it
happens usually in the zeta regularization techniques.

The Laurent series in inverse powers of $\beta$ we obtained is
analogous to the one-loop controbutions to the free energy in
string theory \cite{20}, which is actively investigated in
recent time.

The infinite series (\ref{31a}) and (\ref{32a}) are convergent for
\begin{equation}\beta~<~\beta_C^\pm,~~~\left\{
{\beta_C^-=2\pi/C_s\atop\beta^+_C=\pi/C_{1/2} }\right. .
\label{32}\end{equation}

The series (\ref{31a}) and (\ref{32a}) can be analytically continued
in $\beta$ in the following way.  Using the integral
representation
\[\zeta_R (z)={1\over \Gamma (z)}\int_0^\infty~{t^{z-1}\over{\rm e}^t-1}\]
for the Riemann zeta function and interchanging the orders of
summation and integration, the sums over $k$ can be performed to
yield Bessel functions $J_{k+1}(z)$,
\begin{equation}
\Xi^{\rm -}_{k,2q+1} \left(X\right)
=(-1)^{k+1}~\int_0^\infty{{\rm d}t\over {\rm
e}^t-1}~\left(\left({2\over Xt}\right)^{k+1}{\rm
J}_{k+1}(Xt)-{1\over \Gamma
(k+2)}\right)\,,\label{F1}\end{equation} and Struve functions
$H_{k+3/2}(z)$,
\begin{equation}
\Xi^{\rm -}_{k,2q} \left(X\right)
=(-1)^{k}~\int_0^\infty{{\rm
d}t\over {\rm e}^t-1}~\left({2\over
Xt}\right)^{k+\frac{1}{2}}{\rm H}_{k+\frac{3}{2}}(Xt)\,.
\label{F2}\end{equation}
Taking into account their asymptotic behaviour for large
argument, it is clear that the convergence of the integration
over $t$ breaks down for $x\to \pm {\rm i}$.  For real $X$, the
functions $\Xi^\pm_k(X)$ are smooth. By expanding them back into
a series with respect to powers of $X$, the convergence radius
is given by the nearest pole in the complex $X$-plane which lies
in $X=\pm {\rm i}$ in accordance with (\ref{32}).
For $\beta =\beta_C^-~(X=1)$ and $\beta =\beta^+_C(X=1/2)$ the
functions (\ref{F1}) and (\ref{F2}) remain finite.
{}From the
representations (\ref{F1}) and (\ref{F2}), the behaviour for
$X\to\infty$ can be obtained:
\begin{equation}\Xi^{\rm -}_k \left(X\right)\sim{(-1)^k
\over 2\Gamma (k+2)}~\log X\,,
\label{F1as}\end{equation}
and
\begin{equation}
\Xi^{\rm +}_k \left(X\right)\sim{(-1)^k\over \sqrt{\pi }\Gamma (k+2)}~\log X\,.
\label{F2as}\end{equation}
The functions $\Psi^\pm(X)$ (\ref{psim}) and (\ref{psip}) have similar
properties.
These series are convergent for $X\le 1/(C\pi)$. Taking into account that
their argument is $\beta /2\pi$ (in difference to $C\beta /2\pi$ in case of
the functions $\Xi^\pm$), this is equivalent to (\ref{32}).
These functions can be analytically continued and can be represented in the
form:
\[\Psi^-(X)=\frac{-1}{\pi X}\int_0^\infty{{\rm d}\lambda~\lambda^{2k+1} \over
{\rm e}^{2\pi\lambda}+1}
\log \left({\sinh (\pi X\sqrt{\lambda^2+C^2})
\over \pi X\sqrt{\lambda^2+C^2}  }\right)\]
\[\Psi^+(X)=\frac{-1}{\pi X}\int_0^\infty{{\rm d}\lambda~\lambda^{2k+1} \over
{\rm e}^{2\pi\lambda}-1}
\log \left({\coth (\pi X\sqrt{\lambda^2+C^2})
  }\right).\]
They are of order $O(\frac{1}{X})$ for $X\to\infty$.

\section{Conclusions}
Here we would like to make some final remarks concerning the obtained results.
We have developed a formalism for the calculation of the one-loop free energy
associated with fields of arbitrary spin in the manifold $M=S^1\otimes H^N$.

For the minimally coupled scalar field we have
$C_0^2=\rho_N^2+a^2m^2$, where $\rho_N=(N-1)/2$ and $m$ is the
mass of the field.  For the massless field in $D=4$
dimensions we have $N=3$, $C_0=1$ and $\beta_R =2\pi$ in
agreement with the Rindler temperature $T_R=1/2\pi$, well known in the theory
of conformal invariant fields.
In addition, the leading term of the Laurent series has the form
$-A\pi^2/90\beta^4$, which is also well known \cite{14,21}.
On the other hand the
constant $C_{1/2}$ is the mass of the Dirac spinor field.
For the vector (spin-1) field, the
Hodge-de Rham operator $d\delta+\delta d$ acting on the
exact one-forms is associated with the massless operator
$[-\Delta^\mu\Delta^\nu+(N-1)a^{-2}]g_{\mu\nu}$.
The
eigenvalues of that operator are $\lambda^2+(\rho_N-1)^2$ and for
the Proca-field of mass $m$ we find $C_1^2=(\rho_N-1)^2+a^2m^2$.
In general, for the spin-$s$ field the wave operator has the
form $L^-_N+m^2+Qa^{-2}$, where $Q$ is a given constant.

The renormalized free energy ${\cal F}^-[\overline{g},\beta ]$ of the conformal
invariant scalar field in the ultrastatic space with the metric
$\overline{g}$ is given by Eqs.(\ref{27}) and (\ref{28}). For the conformal
massless field we have $C_0=0$ and hence
\begin{eqnarray}
F^-(\beta )_{2q+1} &=& -\frac{g(0)A(2q+1)}{4\sqrt{\pi}}
\sum_{k=1}^{q}~a^{(0)}_{k,N}\zeta_R(2k+2)\Gamma (k+\frac{1}{2})\Gamma
(k+1)\left({\beta\over 2}\right)^{-2k-2}\,,
\nonumber\\ &&\label{39} \\
F^-(\beta )_{2q} &=& -\frac{g(0)A(2q)}{4\sqrt{\pi}}
\sum_{k=0}^{q-1}~b^{(0)}_{k,N}\zeta_R(2k+3)\Gamma (k+\frac{3}{2})\Gamma
(k+1)\left({\beta\over 2}\right)^{-2k-3}\nonumber \\
&&\hspace*{-2.5cm}+g(0)A(2q)
\sum_{k=0}^{q-1}b_{k,N}^{(0)}\left\{
\left[\frac{\rm d}{{\rm d}z}I^-_k(0,z)_{\mid_{z=0}}+{(-1)^k(1-2^{-2k-1})\over
4(k+1)}\left(\gamma -\log \beta^2 \right)
B_{2k+2}\right]\beta^{-1}\right.\nonumber\\
&&\hspace*{3cm}\left.-2
I^-_k(0,-\frac{1}{2})+\sum_{j=1}^\infty{(-1)^j\over \pi
j}\zeta_R(2j)I^-_k(0,-j)\left({\beta\over
2\pi}\right)^{2j-1}\right\}\,.
\nonumber \\ &&\label{40}\end{eqnarray}
Using eq. (\ref{39}) for $D=4$ dimensions we have $g(0)=1,~A(3)=A/2\pi^2$ and
$F^-(\beta )=-A\pi^2/90\beta^4$.
Thus, there is no term proportional to $\beta^{-2}$, a standard result
\cite{14,21}.
The renormalized stress tenergy ensor can be obtained  by the variation of
${\cal F}^-[g,\beta ]$ with respect to the metric $g$.
This tensor remains finite at the horizon since the divergence of the
thermal contribution obtained by the variation of
${\cal F}^-[\overline{g},\beta ]$
is compensated by the vacuum polarization obtained by  variation of
$\Delta{\cal F}^-[\omega ,g ]$ \cite{21}. In general, for a massive conformal
field the coefficient reads $C_0=(\xi-\frac{1}{6})R+m^2$ \cite{14}.

The dominant contribution to the one-loop entropy density
\begin{equation}
S(\beta )=\hbar^{-1}\beta^2{\partial
F\over\partial\beta}\end{equation}
comes from the region near
the horizon.
In order to regulate the divergencies one can put in an infrared cutoff by
defining a smooth compact manifold $M$ as $M=S^1\otimes H^N/\Gamma$ and them
impose suitable boundary conditions on the fields.
Here, $H^N/\Gamma$ is the quotient of $H^N$ by a discontinuous group $\Gamma$
of isometries.
Under this assumptions
the volume of the compact manifold $H^N/\Gamma$ is $A=V({\cal
F}_N)a^N$, $V({\cal F}_N)$ beeing the volume of the fundamental
domain ${\cal F}_N$.  By making use of the Selberg trace formula
associated with $H^N/\Gamma$ \cite{13}, one obtaines the zeta
function $\zeta(z\in H^N/\Gamma\mid L^\pm_N)=V({\cal
F}_N)\zeta(z\mid L^\pm_N)+\zeta^t(z\mid L^\pm_N)$, where the
topological term (the analytical part of the zeta function)
$\zeta^t(z\mid L^\pm_N)$ can be written as a sum over closed
geodesics of $H^N/\Gamma $. The leading behaviour of the free
energy (\ref{27})-(\ref{30}) is independent on the topological
contributions, of course. So we obtain the same leading behaviour as above with
a finite volume factor.

Moreover, there exist discrete groups $\Gamma$ of special kind
which are related to some noncompact hyperbolic manifolds with
finite volume (in that case the group $\Gamma$ contains
parabolic elements as well) \cite{22}. For the symmetric space of rank-1 some
concrete examples of such groups $\Gamma$ are known.
Therefore the fundamental domain of $\Gamma$ is noncompact and the finite
invariant volume of the domain can be connected with the area of the horizon.

\section*{Acknowledgements}

We would like to thank S. Zerbini for useful discussions.\\
One of the authors (A.A.B.) would be like to thank Leipzig University for kind
hospitality
and the DAAD for a grant.

\end{document}